\documentclass[
    ,final            
  ]
  {aipproc}

\layoutstyle{8x11double}

\begin{document}

\title{
\vspace{-1em}
\begin{flushright}
{\normalsize\rm
KYUSHU-HET-86}
\end{flushright}
Sparticle spectrum and EWSB of
 mixed modulus-anomaly mediation
 in fluxed string compactification models}

\classification{11.25.Mj, 11.25.Wx, 04.65.+e, 12.60.Jv}
\keywords      {flux compactification,
 supersymmetry breaking, moduli, anomaly-mediation}

\author{Ken-ichi Okumura}{
  address={Department of Physics, Kyushu University,
  Fukuoka 812-8581, Japan}
}



\begin{abstract}
We examine low energy sparticle mass-spectrum and electroweak
 symmetry breaking (EWSB) in the mixed modulus-anomaly mediation
 which is naturally realized in KKLT flux compactification.
We find that the anomaly-mediation effectively lowers messenger
 scale in the moduli-mediation and leads to 'squeezed' mass-spectrum
 distinct from any known mediation mechanisms. 
The lightest neutralino typically becomes a mixed state of 
 bino and higgsino or pure higgsino, which has a considerable
 impact on the cold dark matter physics.
\footnote{This talk is based on the work presented in \cite{Choi05}.
}
\end{abstract}

\maketitle


 
\section{Introduction}

In string theory, 
the moduli-mediation \cite{modulimediation}
 has been known for some time as a natural mechanism of supersymmetry (SUSY)
 breaking in low energy 4D effective theory. 
However, any
 reliable predictions can not be derived before all the moduli are
 stabilized and cosmological constant is fine-tuned to the observed vanishing
value \cite{choi3}.

Recently, KKLT have proposed an interesting set-up in which all the moduli
 are stabilized by flux and nonperturbative dynamics on branes \cite{Kachru:2003aw}. Resultant
 SUSY AdS vacuum is uplifted to dS vacuum by introducing
 additional source of SUSY breaking. 
Soft SUSY breaking of visible sector in
 KKLT set-up has been examined carefully
under the assumption that the visible sector
 is sequestered from the SUSY-breaking uplifting sector.
\cite{choi1, choi2}.  It is
 noticed that in this class of models the moduli-mediated SUSY breaking
 is typically ${\it O}(m_{3/2}/4\pi^2 )$
 and the loop-suppressed anomaly-mediation \cite{Randall:1998uk}
 can play equally important roll.

In this paper, we discuss that this mixed modulus-anomaly mediation
 is equivalent to reducing the mediation scale in
 the moduli-mediation up to non-negligible Yukawa interactions and
 realizes distinct pattern of SUSY breaking at low energy
 scale, which leads to its characteristic phenomenology.

\section{KKLT set-up}

Let us first define the KKLT set-up following 
\cite{Kachru:2003aw, choi1, choi2}. 
Throughout this paper, 
we assume $M_{Pl}=1$ otherwise explicitly specified.
We consider type IIB string
 theory compactified on Calabi-Yau (CY) orientifold. Dilaton, $S$ and
 complex structure moduli, $Z^\alpha$ are stabilized by introducing
 NS and RR three form fluxes, while K\"ahler moduli $T^i$ remain
 light at this stage. 
Integrating out $S$, $Z^\alpha$ and assuming a single K\"ahler modulus
$T$, low energy 4D effective action in superconformal formulation of
 N=1 supergravity is given by,
 \begin{eqnarray}
\label{N=1}
S_{N=1}=  
\int d^4x \sqrt{g^C} \,\left[\,
\int d^4\theta \,
CC^*\left(-3\exp\left(-\frac{K_{eff}}{3}\right)\right)
\right.\nonumber \\
+\,\left.\left\{
\int d^2\theta
\left(\frac{1}{4}f_a W^{a\alpha}W^a_\alpha
+C^3W_{eff}\right)
+{\rm h.c.}\right\}\,\right],
\end{eqnarray}
where $C=C_0+\theta^2 F^C$ denotes the chiral compensator superfield.
4D metric in superconformal frame, $g^C_{\mu\nu}$ is related
 to metric in the Einstein frame, $g^E_{\mu\nu}$ by
 $g^C_{\mu\nu}=(CC^\ast)^{-1} e^{K_{eff}/3}g^E_{\mu\nu}$.  
K\"ahler potential and superpotential are given by, 
\begin{eqnarray}
&&K_{eff}=K_0+Z_i(T+T^*)Q_i^*Q_i,
\nonumber \\
&&W_{eff}= W_0+\frac{1}{6}\lambda_{ijk}Q_iQ_jQ_k,
\end{eqnarray}
where $Q_i$ are chiral matter superfields 
while $W^{a\alpha}$ denote chiral field strengths
 of gauge supermultiplet.
They can live on D3 brane or D7 brane wrapping on the 4-cycle
of the CY orientifold. Their kinetic functions are given by,
\begin{eqnarray}
\label{kahlermetric}
Z_i= \frac{1}{(T+T^*)^{n_i}},
&&
f_a=T^{\, l_a},
\end{eqnarray}
where ($n_i=1$, $l_a=0$) for D3 and ($n_i=0$, $l_a=1$) for D7 \cite{ibanez1}.
Matter fields also originate in an intersection of two (three) D7 branes
with $n_i=1/2\,$  ($n_i=1$) \cite{ibanez}. 

\begin{figure}
\includegraphics[height=.3\textheight]{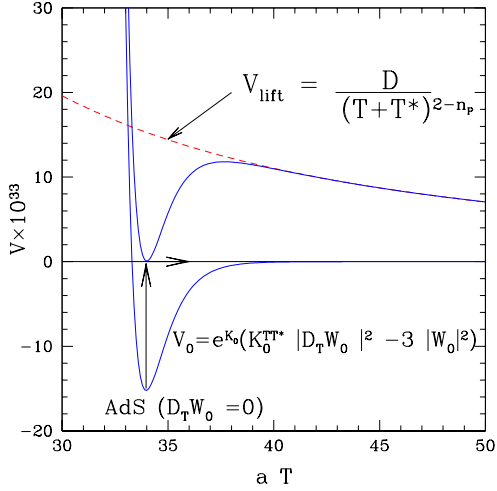}
  \caption{Potential for modulus $T$ in KKLT set-up}
\end{figure}

In the minimal KKLT setup, we introduce a gaugino condensation on D7
 brane to stabilize the K\"ahler modulus $T$. The nonperturbative dynamics generates
 exponential $T$ dependence  in the superpotential and it breaks the no-scale structure of the model:
\begin{eqnarray}
\label{minimal}
K_0= -3\ln(T+T^*), &&
W_0= w_0-Ae^{-aT},
\end{eqnarray}
where $w_0$ is original constant superpotential induced by the fluxes. 
The lower solid curve in Fig. 1 shows potential for $T$ obtained by integrating out
$C$ in the Einstein frame:
\begin{eqnarray}
\label{approx-potential}
&&V_0=e^{K_0}\left(K^{TT^*}_0D_TW_0(D_TW_0)^*-3|W_0|^2\right).
\end{eqnarray}
The potential minimum is given by SUSY AdS vacuum, $D_T W_0=0$
 with $V_0 = -3 |m_{3/2}|^2$
and the K\"ahler modulus is stabilized at
 $\langle aT \rangle \approx \ln\left(A/w_0\right) \approx \ln\left(M_{Pl}/m_{3/2}\right)$.  

KKLT propose to uplift this AdS vacuum to dS vacuum by introducing an explicit 
 source of SUSY breaking.  This is done by $\overline{D3}$ brane
 trapped in a warped throat with a red-shift factor
 $e^{A_{min}}\sim \sqrt{m_{3/2}/M_{Pl}}$ \cite{giddings}. 
In low-energy action, its
 effect is described by a spurion operator up to corrections
 further suppressed by $e^{A_{min}}$ \cite{choi2}:
\begin{eqnarray}
\label{N=0}
&&S_{\rm lift}= 
-\int d^4x \sqrt{g^C} \int d^4\theta \,
\,|C|^4\theta^2
\bar{\theta}^2\,\,{\cal P}_{\rm lift}, 
\label{lifting}
\end{eqnarray}
where ${\cal P}_{\rm lift}=D(T+T^*)^{n_P}$
 with $D\sim e^{4 A_{min}} M_{Pl}^4$
\footnote{In principle, $Q_i$ dependence
 like $X(T+T^*)Q_i Q_i^*$
 can be introduced in (\ref{N=0})
 by the exchange of KK modes besides ${\cal P}_{\rm lift}$, however it is
 expected that the coefficient X is further suppressed by
 the small warp factor, so its contribution to the sfermion mass can be
 ignored compared to the soft masses induced by $F^C$ and $F^T$.}. 
Integrating out C in the Einstein frame, we obtain the following uplifting potential:
\begin{equation}
\label{uplift}
V_{\rm lift}=e^{2K_0/3}{\cal P}_{\rm lift}(T,T^*) \equiv D/(T+T^*)^{2-n_P}.
\end{equation}
Strictly speaking, uplifting by $\overline{D3}$ brane
 predicts constant ${\cal P}_{\rm lift}$ ($n_P=0$), however 
we leave $n_P$ free to cover potential extensions.
Desired dS vacuum can be achieved by fine-tuning $D$ as shown by the upper
solid curve in Fig.1.

 This uplifting process slightly shifts the position of the minimum
 from the original SUSY preserving vacuum and induces non-zero
 $F^T=-e^{K_0/2} K^{TT^*} (D_T W)^*$. While $m_{3/2}=e^{K_0/2} W_0$
 is almost constant and $V_0/3=-|m_{3/2}|^2$ is canceled
by the uplifting. This leads to a hierarchy:
\begin{equation}
\left|\frac{F^T}{(T+T^\ast)}\right|^2 = \frac{K_{TT^\ast}}{3}|F^T|^2 \approx |\delta m_{3/2}^2| <<
 |m_{3/2}|^2.
\end{equation} 
Actual minimization of the uplifted potential gives,
\begin{eqnarray}
\label{C/T}
\frac{F^C}{C_0} &\approx& m_{3/2} \,\approx\, \frac{w_0}{M_{Pl}^2(T+T^*)^{3/2}},
\nonumber \\
\frac{F^T}{(T+T^*)} &\approx& \frac{2-n_P}{a(T+T^*)}m_{3/2},\,
\,\approx\, {\cal O}\left(\frac{m_{3/2}}{4\pi^2}\right),
\end{eqnarray}
suggesting that the loop-suppressed anomaly mediation is
 comparable to the moduli-mediation \cite{choi1, choi2}.
Here, we define $\alpha$ which parameterizes their relative significance:
\begin{equation}
\label{alpha}
\alpha \equiv \frac{m_{3/2}/\ln(M_{Pl}/m_{3/2})}{M_0}\,\approx\,
\frac{F^C/C_0}{a {\rm Re}(T)}
\left(\frac{F^T}{(T+T^*)}\right)^{-1}.
\end{equation}

\section{Mirage messenger scale}

We calculate low energy mass-spectrum of visible fields in 
 KKLT set-up. Their soft SUSY breaking is written as
\footnote{Gaugino vertex is given by, ${\cal L} = i\sqrt{2} \tilde{Q}_i
 T^a q_i \lambda^a + {\rm h.c.}$ where $q_i$ denote fermions and $T^a$ is a
 generator of gauge group.}
,
\begin{equation}
{\cal L}_{soft}=-m_i^2|\tilde{Q}_i|^2
-\left(\frac{1}{2}M_a\lambda^a\lambda^a+\frac{1}{6}A_{ijk}y_{ijk}\tilde{Q}_i\tilde{Q}_j\tilde{Q}_k +{\rm h.c.}
\right),
\end{equation}
where $\lambda^a$ denote gauginos, $\tilde{Q}_i$ represent sfermions and
$y_{ijk}$ are canonically normalized Yukawa couplings. 
Just below the cut-off,
 the standard calculation \cite{modulimediation, Randall:1998uk} gives,
\begin{eqnarray}
\label{soft1}
M_a
&=& l_a M_0+\frac{b_a}{8\pi^2}g^2_{GUT}\frac{F^C}{C_0},
\nonumber \\
A_{ijk}
&=&a_{ijk}M_0
-\frac{1}{16\pi^2}(\gamma_i+\gamma_j+\gamma_k)\frac{F^C}{C_0},
\nonumber \\
m_i^2
&=&c_i|M_0|^2
-\frac{1}{32\pi^2}\frac{d\gamma_i}{d\ln\mu}\left|\frac{F^C}{C_0}\right|^2
\nonumber\\
&&
+\frac{1}{8\pi^2}
\left\{
\sum_{jk}a_{ijk}\left|\frac{y_{ijk}}{2}\right|^2-
\sum_A l_Ag_A^2C_A(Q_i)\right\}
\nonumber\\
&&
\phantom{+\frac{1}{8\pi^2}}\times
\left(M_0\left(\frac{F^C}{C_0}\right)^*
+M_0^* \left(\frac{F^C}{C_0}\right)\right),
\end{eqnarray}
where $M_0 \equiv F^T/(T+T^\ast)$, $a_{ijk}=3-n_i-n_j-n_k$
 and $c_i=1-n_i$ for the moduli-mediated contribution and, 
\begin{eqnarray}
\frac{dg_a}{d\ln \mu}=\frac{b_a}{8\pi^2} g_a^3, &&
 \frac{d\ln Z_i}{d\ln \mu}=\frac{1}{8\pi^2}\gamma_i,
\end{eqnarray}
for the anomaly-mediated contribution.
Note that interference terms appear due to non-zero $\partial_T \gamma_i$
\cite{choi1,choi2}.
If visible gauge fields originate in D3 and no Yukawa coupling
 with D7 singlets, the above formulas reduce to the pure anomaly-mediation.
Because it has been examined extensively
 by the literature and is plagued with tachyonic slepton, we 
 assume D7 visible gauge fields ($l_a=1$)
 which naturally lead to gauge coupling unification.   

Surprising new feature of this mixed modulus-anomaly mediation
 arises from correspondence between
 renormalization group (RG) running in the moduli-mediation and 
 the anomaly-mediation itself.
They have precisely same form at one-loop if $a_{ijk}=1$ and
 Yukawa couplings are vanishing except for fields
 satisfying $c_i+c_j+c_k=1$. Thus we can obtain analytic
 solutions for the soft terms in the mixed modulus-anomaly
 mediation up to corrections coming from Yukawa couplings
 with $n_i+n_j+n_k \neq 2$:
\begin{eqnarray}
M_a(\mu)&=&
M_0 
-\frac{M_0}{4\pi^2}b_ag_a^2(\mu)\ln\left(\frac{M_{MMS}}{\mu}\right)
,\\
A_{ijk}(\mu)&=&a_{ijk}\, M_0 
+\frac{M_0}{8\pi^2}(\gamma_i(\mu)
+\gamma_j(\mu)+\gamma_k(\mu))
\nonumber\\
&&
\phantom{a_{ijk}\,M_0+}
\times 
\ln\left(\frac{M_{MMS}}{\mu}\right),
\\
m_i^2(\mu)&=&  
c_i 
\left|M_0\right|^2 
\nonumber\\
&&~
+\frac{\left|M_0\right|^2}{4\pi^2}
\left\{\gamma_i(\mu)
-\frac{1}{2}\frac{d\gamma_i(\mu)}{d\ln\mu}
\ln\left(\frac{M_{MMS}}{\mu}\right)\right\}
\nonumber\\
&&~\phantom{\frac{\left|M_0\right|^2}{4\pi^2}}
\times
\ln\left(\frac{M_{MMS}}{\mu}\right),
\end{eqnarray}
where $M_{MMS}\equiv \left(m_{3/2}/M_{Pl}\right)^{\alpha/2} M_{\rm GUT}$ and we used $F^C/C_0 \approx m_{3/2}$ and (\ref{alpha}) 
\footnote{If $Q_i$ have $U(1)$ charges $Y_i$ and $Tr(cY)$ is non-zero, $Y_i$ components in $c_i |M_0|^2$ behave as
an effective  D-term. Thus further corrections  appear for these
components due to
 running of the $U(1)$ coupling.}.
These solutions
show that net effect of the anomaly mediation is shifting messenger
scale in the moduli-mediation from $M_{GUT}$ to {\it the mirage messenger
scale}, $M_{MMS}$. It is noted that this scale is not associated with
any physical threshold and cut-off of the theory still stays at $M_{GUT}$.
This new feature of the
anomaly-mediation does not depend on detail of the co-existing
 mediation scheme as far as the above conditions are satisfied.


\begin{figure}
  \includegraphics[height=.3\textheight]{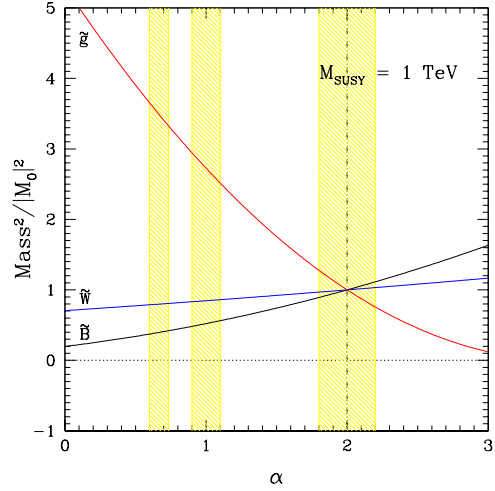}
  \caption{Gaugino mass squares in KKLT set-up (D7 visible
 gauge fields).}
\end{figure}
\begin{figure}
  \includegraphics[height=.3\textheight]{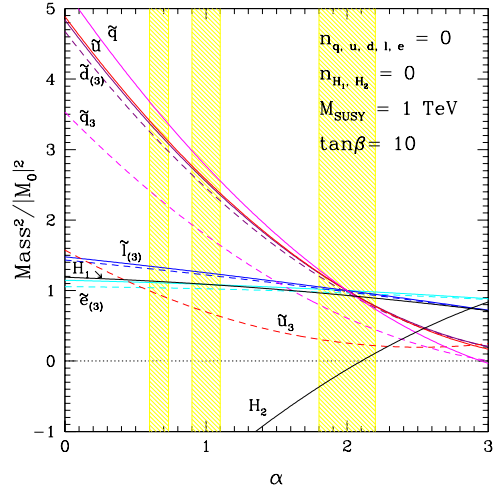}
  \caption{Scalar mass squares in KKLT set-up (D7 visible
 gauge/matter fields).
}
\end{figure}

\section{Low energy phenomenology}
Fig. 2, 3 show the results of numerical calculation for
 the gaugino and sfermion mass squares at $M_{SUSY}=1$ TeV. 
Gauge and Yukawa couplings at $M_{SUSY}$ are estimated
 using 2-loop RG equations in the SM and subsequently evolved to
 $M_{GUT}=2\times 10^{16}$ GeV by 1-loop RG equations in the MSSM. 
The soft SUSY breaking terms are calculated by solving
 1-loop RG equations imposing the boundary conditions (\ref{soft1}) at $M_{GUT}$. 
We choose all the visible fields originating in D7 ($n_i=0$).
Shaded regions at $\alpha=2/3, 1, 2$ correspond to the cases, $n_P=-1,0,1$
 assuming $10\%$ ambiguity. The minimal KKLT predicts
 $\alpha_{KKLT}\approx 1$ and $M_{MMS}\approx 10^{10}$ GeV. 
Reduced mediation scale generally leads to squeezed (compressed)
 mass spectrum, which is distinct from any known mediation mechanisms. 
In particular, 
 we have unified gaugino
 and the 1st and 2nd generation sfermion masses at $\alpha=2$
 because of $M_{MMS}\approx M_{SUSY}$.
Higgs and the third generation masses (dashed curves) are
 disturbed by Yukawa couplings.
Note that these predictions are different from those of
 lower cut-off models
 because here gauge couplings still unify at $M_{GUT}$.
These distinct pattern of the spectrum can be tested
 by coming LHC and future ILC.

%

Let us examine the electroweak (EW) symmetry breaking in the
 mixed modulus-anomaly mediation.
Here we assume the minimal model with $\mu$
 and $B$ terms:
\begin{eqnarray}
W_H = \mu H_1 H_2, && {\cal L}_{soft} = B\mu H_1 H_2. 
\end{eqnarray}
They can be solved by minimizing the tree-level Higgs potential
 for fixed $\tan\beta=\langle H_2 \rangle/\langle H_1 \rangle$, 
\begin{eqnarray}
\label{eq:rewsb_condition}
\mu^2 &=& -\frac{M_Z^2}{2}
+\frac{m^2_{H_1}-m^2_{H_2}\tan^2\beta}{\tan^2\beta-1}, \nonumber\\
|B\mu| &=& \sin(2\beta)(m^2_{H_1}+m^2_{H_2}+2\mu^2)/2.
\end{eqnarray}
Figure 4. shows $\mu$ and $B$ in the unit of $M_0$.
Here all the visible matters live in the intersection
 ($n_i=1/2$) and we fix $\tan\beta=10$.
The dashed (thin solid) curves indicate the case, $M_Z = 0.3 M_0$ ($M_Z << M_0$). 
$M_Z^2>0$ in (\ref{eq:rewsb_condition}) ensures that there's no EW symmetry
breaking above the thin solid $|\mu/M_0|$ curve.
Increasing $\alpha$, lowered messenger scale weakens the radiative EW
symmetry breaking \cite{rewsb} 
and reduces $|\mu|$. Eventually $|\mu|^2$
is driven into negative and EW symmetry is restored. 
While the bino mass increases with $\alpha$, again because of the reduced 
messenger scale. These effects drive the lightest neutralino
 into a mixed state of bino and Higgsino ($\alpha \approx 1$) 
or pure Higgsino ($\alpha \approx 2$). This qualitative feature
 is robust for different choices of matter fields although
 precise position of the mixed region and nature of the lightest
 supersymmetric particle (LSP) depend on the models. 
Generally, heavier initial
 Higgs mass or lower $M_0/M_Z$ tends to introduce stop LSP and D3 slepton
 or large $\tan\beta$ likely leads to stau LSP. If they are stable,
they will pose severe constraints on the models.
If the mixed or pure Higgsino neutralino is LSP,
 faster annihilation rate and stronger interaction with nuclei
 will dramatically
 change the cold dark matter physics based on the conventional
 bino LSP scenario \cite{darkmatter}. 
Thus the mixed modulus-anomaly mediation
 can have a considerable impact on cosmology as well as with hierarchically
 heavy gravitino and moduli fields \cite{Kohri:2005ru, Endo:2005uy}.


\begin{figure}
  \includegraphics[height=.3\textheight]{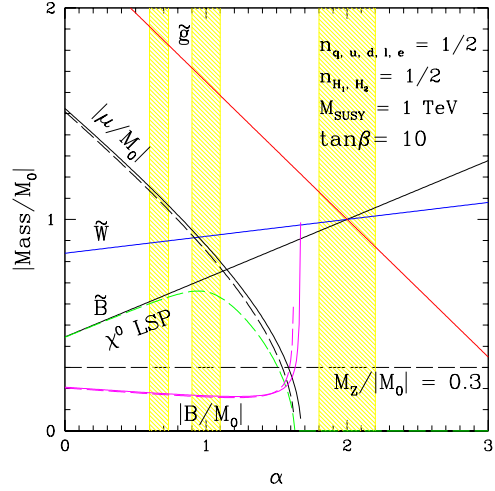}
  \caption{Electroweak symmetry breaking in KKLT set-up. 
Gauge fields originate in $D7$ while
all the matter
 fields originate in an intersection of $D7$.
The dashed (thin solid) curves correspond to $M_Z/M_0 = 0.3$ ($M_Z/M_0 \approx 0$).}
\end{figure}


\begin{theacknowledgments}

This work is supported
 by KRF PBRG 2002-070-C00022, the BK21 program of Ministry
 of Education and the Center for High Energy Physics of
Kyungbook National University  and in part
 by  the Grand-in-aid for Scientific Research on
Priority Areas \#441: ``Progress in elementary particle physics of the
21st century through discovery of Higgs boson and supersymmetry''
 \#16081209
from the Ministry of Education, Culture,
Sports, Science and Technology of Japan.

\end{theacknowledgments}



%
%

\end{document}